\documentclass[conference]{IEEEtran}
\IEEEoverridecommandlockouts

\usepackage{cite}
\usepackage{amsmath,amssymb,amsfonts}
\usepackage{algorithmic}
\usepackage{graphicx}
\usepackage{textcomp}
\usepackage{xcolor}
\usepackage{multirow}
\usepackage{multicol}
\usepackage[colorlinks,bookmarksopen,bookmarksnumbered,citecolor=green, linkcolor=red, urlcolor=green]{hyperref}
\def\BibTeX{{\rm B\kern-.05em{\sc i\kern-.025em b}\kern-.08em
    T\kern-.1667em\lower.7ex\hbox{E}\kern-.125emX}}
\begin{document}

\title{Beyond Point Annotation: A Weakly Supervised Network Guided by Multi-Level Labels Generated from Four-Point Annotation for Thyroid Nodule Segmentation in Ultrasound Image}


\author{
\IEEEauthorblockN{Jianning Chi\textsuperscript{1}, Zelan Li\textsuperscript{1}, Huixuan Wu\textsuperscript{1}, Wenjun Zhang\textsuperscript{2}, Ying Huang\textsuperscript{3}}
\IEEEauthorblockA{\textsuperscript{1}\textit{Faculty of Robot Science and Engineering, Northeastern University, Shenyang, China}}
\IEEEauthorblockA{\textsuperscript{2}\textit{Division of Biomedical Engineering, University of Saskatchewan, Saskatoon, Canada}}
\IEEEauthorblockA{\textsuperscript{3}\textit{Department of Ultrasound, Shengjing Hospital of China Medical University, Shenyang, China}}
}

\maketitle

\begin{abstract}
Weakly-supervised methods typically guided the pixel-wise training by comparing the predictions to single-level labels containing diverse segmentation-related information at once, but struggled to represent delicate feature differences between nodule and background regions and confused incorrect information, resulting in underfitting or overfitting in the segmentation predictions. In this work, we propose a weakly-supervised network that generates multi-level labels from four-point annotation to refine diverse constraints for delicate nodule segmentation. The Distance-Similarity Fusion Prior referring to the points annotations filters out information irrelevant to nodules. The bounding box and pure foreground/background labels, generated from the point annotation, guarantee the rationality of the prediction in the arrangement of target localization and the spatial distribution of target/background regions, respectively. Our proposed network outperforms existing weakly-supervised methods on two public datasets with respect to the accuracy and robustness, improving the applicability of deep-learning based segmentation in the clinical practice of thyroid nodule diagnosis.


\end{abstract}

\begin{IEEEkeywords}
Medical image processing, ultrasound nodule segmentation, weakly supervision segmentation, deep learning.
\end{IEEEkeywords}

\section{Introduction}
Thyroid diseases have become increasingly prevalent with an annual incidence rate exceeding 10\% in the past decade~\cite{chi2023htunet}. The American College of Radiology’s Thyroid Imaging Reporting and Data System (TI-RADS) reveals that the morphology and borders of thyroid nodules are important ultrasound features for diagnosing malignant nodules~\cite{tessler2017Ti-rads}. Therefore, the segmentation of the thyroid nodules in the ultrasound image plays a crucial role in the diagnosis of thyroid diseases in the early stage. 

Recently, deep learning algorithms have been extensively studied to tackle segmentation problems of thyroid nodules. However, most of these methods claim that convincing segmentation performance relies on large number of precisely ground-truth mask~\cite{ronneberger2015unet,cai2020denseunet,tao2022cenet}, which is labor-intensive and time-consuming. In contrast, weakly-supervised networks have gained increasing attention by use of coarse labels such as box labels~\cite{zhang2020scrf,mahani2022uncrf,tian2021BoxInst} and point labels~\cite{li2023wsdac,zhao2024IDMPS}, offer time and resource efficiency. For example, Zhang et al.~\cite{zhang2020scrf} enforced a local label consistency by spatially constrained to generate Single pixel pseudo label. Mahani et al.~\cite{mahani2022uncrf} employed box labels supervised by an uncertainty-guided cross-entropy loss to improve segmentation accuracy. However, these method used confused multiple information brought difficulties for networks to represent delicate feature differences between nodule and background regions. Li~\cite{li2023wsdac} and Zhao~\cite{zhao2024IDMPS} utilized the point annotations to generate the pairwise pseudo-labels for thyroid segmentation, but inevitably introduced erroneous information and mislead the learning of nodule shapes. Tian et al.~\cite{tian2021BoxInst} redesigned the loss function for instance segmentation based on rectangular box labels, inspiring the design of our network, but the color based pairwise loss was not suitable for blurry ultrasound images to determine boundaries.


In this paper, we propose a weakly-supervised network guided by multi-level labels generated from four-point annotation named as $\text{PA}^+$Net. This network introduces multiple constraints derived from the point annotation to refine the input-level and prediction-level information for segmenting nodules with accurate and delicate boundaries. Specifically, the Distance-Similarity Fusion Prior, based on the spatial distance and grayscale difference between image pixels and annotated points, filters the input image information with low relevance to nodules. While for the network training, the bounding box label generated from the point annotation guarantees the prediction from the segmentation head correctly located in the arrangement of the nodule, and the pure foreground/background labels guide the features projected from the backbone network by the projection head rationally partitioned as nodule or background in both pixel-scale and patch-scale image domain. By refining multiple information into multi-level labels, our proposed network learns features under multi-scale constraints and outperforms other weakly supervised networks, with Jaccard and Dice coefficients of 82.36\% and 89.85\% on the TN3K dataset~\cite{gong2021tn3k} and 84.33\% and 91.37\% on the DDTI dataset~\cite{pedraza2015DDTI}. The code will be made public on \href{https://github.com/bluehenglee/PAplusNet}{https://github.com/bluehenglee/PAplusNet}. 

The contributions of the proposed method are summarized as follows:
\begin{enumerate}
    \item To address the issue of confusing information provided by weak supervision labels, we refine diverse constraints from points labels, introducing a Distance-Similarity Fusion Prior for nodule location, and box labels as well as pure foreground/background labels for spatial distribution.
    \item To address the problem of error messages introduced by single pixel-wise pseudo-labels, we decouple the network supervision into multi-scale constraints. The region-scale alignment loss used to compare predicted minimum bounding box with box labels to locate nodule target. The pixel-scale and patch-scale contrastive loss used to learn distinguish feature from pure foreground/background labels.
\end{enumerate}

\section{Proposed Method}
\begin{figure*}[htbp!]
\centering
\includegraphics[width=0.9\textwidth]{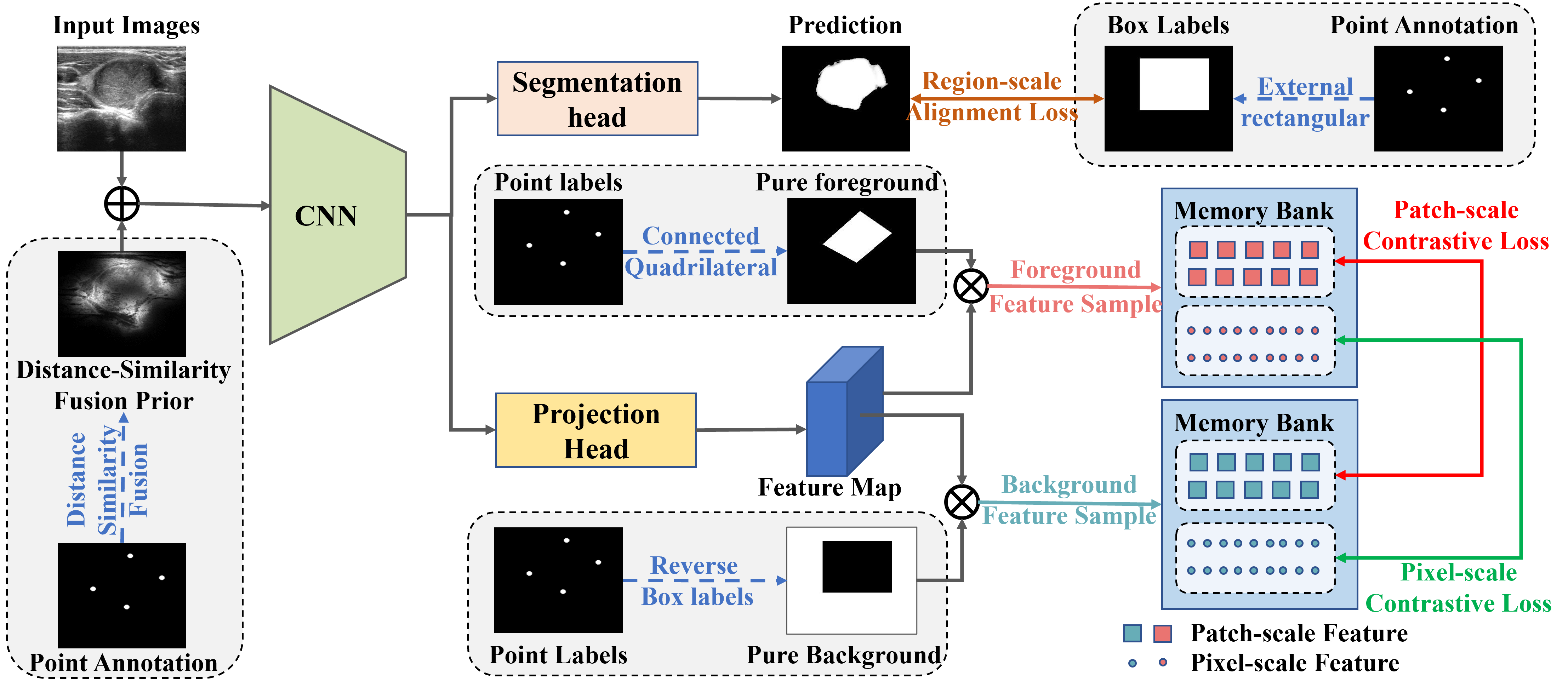}
\caption{\textbf{The overall architecture of proposed $\text{PA}^+$Net.} Images and the Distance-Similarity Fusion Prior are concatenated and fed into the backbone network. The box label is compared with the segmentation predictions using the region-scale alignment loss. The contrastive loss function guides the features projected from the projection head to be rationally partitioned into nodule or background categories at both pixel and patch scales, based on pure foreground/background labels.
}
\vspace{-0.2in}
\label{pic: network overview}
\end{figure*}

In our proposed network, the four-point annotation is used to extract multi-level information. The Distance-Similarity Fusion Prior is concatenated with input images to guide the network's focus around the target area. The box label helps the network learn target locations by adjusting the predictions according to the region-level alignment loss. Pure foreground/background labels enable the network to learn discriminative features from the pure foreground and background areas, without being affected by incorrect information, allowing for delicate delineation of nodule boundaries.


\subsection{Multi-level Information Extraction}
\subsubsection{Distance-Similarity Fusion Prior}
Although pixels between different organs exhibit high similarity in low-contrast medical images, identical organs typically occupy contiguous spaces due to the anatomical spatial relationships within the human body. The Distance-Similarity Fusion Prior quantifies the spatial and feature similarity between points and each pixel in the image. It concentrates on the regions around nodule according to the Euclidean distance and provides edge response for segmentation according to the pixel similarity.

Given one annotated point $(x_i,y_i)$ and a pixel $(x_j,y_j)$ in the image, the Distance-Similarity Fusion Prior is obtained as the following formula:
\begin{eqnarray}
d_{ij} = \frac{\sqrt{(x_i-x_j)^2 + (y_i-y_j)^2}}{\sqrt{w^2+h^2}}
\label{eq:dij}
\end{eqnarray}
\begin{eqnarray}
s_{{ij}} = |{I}\left({x}_{{i}}, {y}_{{i}}\right)-{I}\left({x}_{{j}}, {y}_{{j}}\right)|
\label{eq:sij}
\end{eqnarray}
where $d_{ij}$ represents the normalized distance between the $i$-th point and the $j$-th image pixels, and $s_{{ij}}$ represent the similarity map. ${I}\left({x}_{{k}}, {y}_{{k}}\right)$ represents the pixel value of the image at coordinates $(x_k,y_k)$, and $|\   |$ represents the absolute value. The weighted distance map ${w}_{{d}}$ and similarity map ${w}_{{s}}$ are then defined as follows:
\begin{equation}
{w}_{{d}} = \left\{
\begin{array}{ll}
1, & {x}_{i} = {x}_{j} \text{ and } {y}_{i} = {y}_{j} \\
{e}^{-\frac{{d}_{j}^{2}}{2 \sigma^{2}}}, & \text{otherwise}
\end{array}
\right.
\label{eq:wd}
\end{equation}
\begin{equation}
{w}_{{s}} = \left\{
\begin{array}{ll}
1, & {x}_{i} = {x}_{j} \ \text{and} \  {y}_{i} = {y}_{j} \\
{e}^{-\frac{{s}_{ij}}{2 \theta}}, & \text{otherwise}
\end{array}
\right.
\label{eq:ws}
\end{equation}
where $\sigma$ is the standard deviation of the Gaussian distribution and $\theta$ represents the standard deviation of the similarity coefficient. As $\sigma$ increases, the range of the Prior focus becomes larger. A higher value of $\theta$ implies a stricter requirement for pixel similarity. Finally, the fused Distance-Similarity Fusion Prior $D_{ij}$ can be expressed as:
\begin{eqnarray}
D_{ij} = w_{d_{ij}} \cdot w_{s_{ij}}.
\end{eqnarray}

\subsubsection{Box Label and Pure foreground/background Labels Generation}

As shown in Fig.~\ref{pic: network overview}, by performing a series of geometry operations on the point annotation drawn by the doctors in the clinical examination, we can obtain multi-level labels, including the box label and pure foreground/background labels, which contains different information about the nodule segmentation.
\begin{table}[htbp]
    \centering
    \caption{The pixel classification accuracy of different labels. Pure f/b labels represent pure foreground/background labels.}
        \begin{tabular}{l|c|c|c|c}
            \hline
            \multirow{2}{*}{labels precision} & \multicolumn{2}{c|}{TN3K} & \multicolumn{2}{c}{DDTI} \\
            \cline{2-5}
            & foreground & background & foreground & background \\
            \hline
            pure f/b labels & 98.17\% & 99.98\% & 99.65\% & 99.98\%\\
            box labels & 66.14\% & 99.98\% & 73.64\% & 99.98\% \\
            BoxInst labels & 47.17\% & 95.69\% & 61.97\% & 96.85\% \\
            ellipse labels & 82.97\% & 98.45\% & 89.13\% & 99.80\% \\
            \hline
        \end{tabular}
    \label{tab:ablation_results}
\end{table}

For the box label generation, we identify the minimum bounding rectangle formed by point labels, the box labels include all foreground and some mixed-background pixels, can indicate the thyroid location. 

For the foreground/background labels generation, by connecting point annotation in a sequence, a quadrilateral region can be generated and filled to serve as the pure foreground label. Negating the box label yields the pure background label containing only background pixels. The precision of pure domain labels exceeds 98\% in both the foreground and background. The average precision for box labels used in SCRF~\cite{zhang2020scrf} and UNSCRF~\cite{mahani2022uncrf} algorithm in the foreground is approximately 70\%. The BoxInst labels generated from BoxInst algorithm~\cite{tian2021BoxInst} achieve nearly 60\% average precision. And Ellipse labels~\cite{zhao2024IDMPS} in IDMPS~\cite{zhao2024IDMPS} reach average precision around 85\%. Our multi-level labels, used in network training, are more reliable and accurate than other pseudo-labels in both the foreground and background areas.

\subsection{Multi-scale Constraints}
According to the above-mentioned box label and pure foreground/background labels, we propose two loss functions to guide the network to learn accurate prediction and refining boundaries. The region-scale alignment loss helps the network correctly locate nodules based on box labels. The pixel-scale and patch-scale contrastive loss functions enhance feature representation to refine the segmentation boundaries.

\subsubsection{Region-scale Alignment Loss for Nodule Location}
For nodule localization, the predicted result should be spatially aligned with the ground truth. Without considering the shape of the segmentation, we can represent the alignment condition as the overlap between the external bounding box regions of the ground truth and the predicted result. Specifically, for the prediction $M_i$ and the box label $Y_{i}^{box}$ generated from points annotation. The alignment loss $L_{a}$ can be defined as:
\begin{eqnarray}
L_{a} = Dice(M_{i}^{box}, Y_{i}^{box})
\end{eqnarray}
where $M_{i}^{box}$ denote the bounding box regions generated by the predicted result. Alignment loss guides the network to learn target location by only using the position information provided by the box label, avoiding the noisy information introduced by pixel-wise comparison between the prediction and the box label.


\subsubsection{Pixel-scale and Patch-scale Contrastive Loss for Nodule Shape Refinement}
By using pure foreground/background labels, the thyroid image can be partitioned into \textbf{pure} foreground, \textbf{pure} background, and \textbf{mixed} areas. The multi-scale contrastive loss learn discriminative feature expressions from pure areas to further segment the mixed area. Importantly, the sampled pairs are drawn from the entire training batch, rather than the same image. The generic contrastive loss with an m-scale sampling size, denoted as $L_{c}^m$, is defined as follows:
\begin{eqnarray}
L_{c}^m = \frac{1}{n} \sum_{q^{+} \in P_{i}} -\log \frac{e^{\frac{q \cdot q^{+}}{\tau}}}{e^{\frac{q \cdot q^{+}}{\tau}} + \sum_{q^{-} \in N_{i}} e^{\frac{q \cdot q^{-}}{\tau}}},
\end{eqnarray}
where $P_{i}$ and $N_{i}$ denote the sets of positive and negative feature embedding queues for iteration $i$, respectively. $q^{+}$ represents a foreground feature sample, and $q^{-}$ denotes a sample in background feature queues, while $q$ is the anchor feature queues selected from the foreground. $\tau > 0$ is a temperature parameter that controls the slope of the loss function and affects its smoothness. $\cdot$ denotes the inner product operation. 
The contrastive loss function makes features of the same category closer in the embedding space, while features of different categories are pushed further apart. When calculating the contrastive loss at pixel scale ($m=1$), the pairs used are sampled pixel features. Similarly, based on experimental results, we calculate the patch-level contrastive loss at scale size ($m=3$), where the pairs used are patch features.



\subsubsection{Loss Function}
The overall loss function consists of alignment loss for target location and contrastive loss for pixel classification. The alignment loss $L_{a}$ supervision from a region-scale, and the contrastive loss uses a pixel-scale $L_{c}^1$ and patch-scale contrast $L_{c}^3$ learning at the same time. Based on the above, the overall loss function of the network can be expressed as:
\begin{eqnarray}
L_{all} = L_{a} + L_{c}^1 + L_{c}^3
\end{eqnarray}


\section{Experimental Results}
\subsection{Dataset}
The public dataset TN3K includes 3,494 images, with 2,879 images in the training set and 614 images in the test set. The DDTI dataset contains 637 thyroid nodule images, and we apply 10-fold cross-validation for DDTI.

We conducted a quantitative comparison using three common segmentation evaluation metrics: Mean Intersection over Union (mIoU) and Dice similarity coefficient (DSC) for evaluating segmentation accuracy, and Hausdorff distance (HD) for assessing segmentation boundary precision.

\begin{table}[t!]
    \centering
    \caption{Quantitative results of Models C on TN3K and DDTI Datasets with different consistency loss weight. DS mean distance similarity. The units of miou and DSC are (\%), and the unit of Hd is (mm).}
        \begin{tabular}{c|c|cc|cc}
            \hline
            \multirow{2}{*}{Model} & \multirow{2}{*}{Strategy} & \multicolumn{2}{c|}{TN3K} & \multicolumn{2}{c}{DDTI} \\
            \cline{3-6}
            & & DSC$\uparrow$ & HD$\downarrow$ & DSC$\uparrow$ & HD$\downarrow$ \\
            \hline
            A & Baseline & 67.65 & 6.44 & 59.68 & 7.68\\
            B & +DS & 79.76 & 5.78 & 83.49 & 6.75 \\
            C & +$L_{a}$ & 72.16 & 5.79 & 65.28 & 7.43 \\
            D & +$L_{c}$ & 76.91 & 12.96 & 69.34 & 12.22 \\
            E & +$L_{a}$+$L_{c}$ & 79.11 & 5.13 & 72.14 & 6.67\\
            F & +DS+$L_{a}$ & 84.86 & 4.68 & 88.35 & 5.02 \\
            G & +DS+$L_{c}$ & 83.44  & 8.58 & 84.07 & 11.79 \\
            H & +DS+$L_{a}$+$L_{c}$ & \textbf{89.85} & \textbf{3.89} & \textbf{91.37} & \textbf{4.35}\\
            \hline
        \end{tabular}
\label{tab:ablation_results}
\vspace{-0.15in}
\end{table}

\begin{table}[htbp!]
\centering
\caption{Quantitative comparison results of different methods on the TN3K and DDTI datasets. The units of miou and DSC are (\%), and the unit of Hd is (mm)}
\begin{tabular}{c|ccc|ccc}
\hline
\multirow{2}{*}{Method} & \multicolumn{3}{c|}{TN3K} & \multicolumn{3}{c}{DDTI} \\
\cline{2-7}
 & mIoU$\uparrow$ & HD$\downarrow$ & DSC$\uparrow$ & mIoU$\uparrow$ & HD$\downarrow$ & DSC$\uparrow$ \\
\hline
U-Net\cite{ronneberger2015unet} & \textbf{83.53} & 3.94 & \textbf{90.52} & 89.27 & 4.26 & 94.21 \\
Denseunet\cite{cai2020denseunet} & 80.86 & 3.94 & 88.70 & 87.67 & 4.66 & 93.29 \\
Cenet\cite{tao2022cenet} & 83.46 & \textbf{3.84} & 90.47 & \textbf{93.74} & \textbf{3.27} & \textbf{96.66} \\
\hline
SCRF\cite{zhang2020scrf} & 69.83 & 6.06 & 81.64 & 71.96 & 6.73 & 83.30 \\
UNCRF\cite{mahani2022uncrf} & 74.45  & 5.63 & 84.82 & 77.31 & 5.98 & 87.04 \\
BoxInst\cite{tian2021BoxInst} & 79.92 & 4.31 & 88.21 & 81.81 & 4.64 & 89.74 \\
WSDAC\cite{li2023wsdac} & 68.65 & 4.49 & 80.65 & 74.02 & 4.61 & 84.91 \\
IDMPS\cite{zhao2024IDMPS} & 72.08 & 4.06 & 83.25 & 83.46 & 4.56 & 90.41 \\
Proposed & \textbf{82.36} & \textbf{3.89} & \textbf{89.85} & \textbf{84.33} & \textbf{4.35} & \textbf{91.37} \\
\hline
\end{tabular}
\vspace{-0.15in}
\label{tab:comparison_result}
\end{table}

\subsection{Ablation Experiments}

\subsubsection{Effectiveness of Distance-Similarity Fusion Prior}
As shown in Table~\ref{tab:ablation_results}, by comparing models A with B, C with F, D with G, and E with H, it can be considered that Distance-Similarity Fusion Prior led to enhancements in the segmentation results. Depending on the combination, the Dice Similarity Coefficient (DSC) increased by 6.53\% to 12.11\%, and the Hausdorff Distance (HD) decreased from 4.38 to 0.66, demonstrating the effectiveness of the Distance-Similarity Fusion Prior.


\subsubsection{Effectiveness of Alignment Loss}
The introduction of alignment loss significantly improves segmentation performance on two thyroid nodule datasets. As detailed in Table~\ref{tab:ablation_results}, comparisons between models A and C, B and F, and G and H show that alignment loss enabled the network to learn more accurate locations, leading to approximately a 5\% improvement in the average DSC. More importantly, alignment loss plays a crucial role in enhancing contrastive loss performance. When comparing models D with E, the Hausdorff distance was reduced from 12.96 to 5.13 in the TN3K dataset, and from 12.22 to 6.67 in the DDTI dataset.


\subsubsection{Effectiveness of Contrastive Loss}
Table~\ref{tab:ablation_results} shows that integrating contrastive loss significantly improved the shape and edge definition of segmentation contours. This improvement was reflected in the DSC scores, which increased by 9.26\% in the TN3K dataset and by 9.66\% in the DDTI dataset when comparing model A to model D. Furthermore, comparing models C and E, the HD values dropped by 0.63 in TN3K and 0.73 in DDTI when alignment loss was applied together.


\subsection{Comparison with State-of-the-art Method}
We compare the proposed $\text{PA}^+$Net with classical fully supervised networks and other state-of-art weakly supervised segmentation methods, including U-Net, Densenet, Cenet, SCRF, UNCRF, BoxInst, WSDAC, IDMPS. During validation, the point annotation was provided to all networks.

\begin{figure}[!htbp]
\centering
\includegraphics[width=0.48\textwidth]{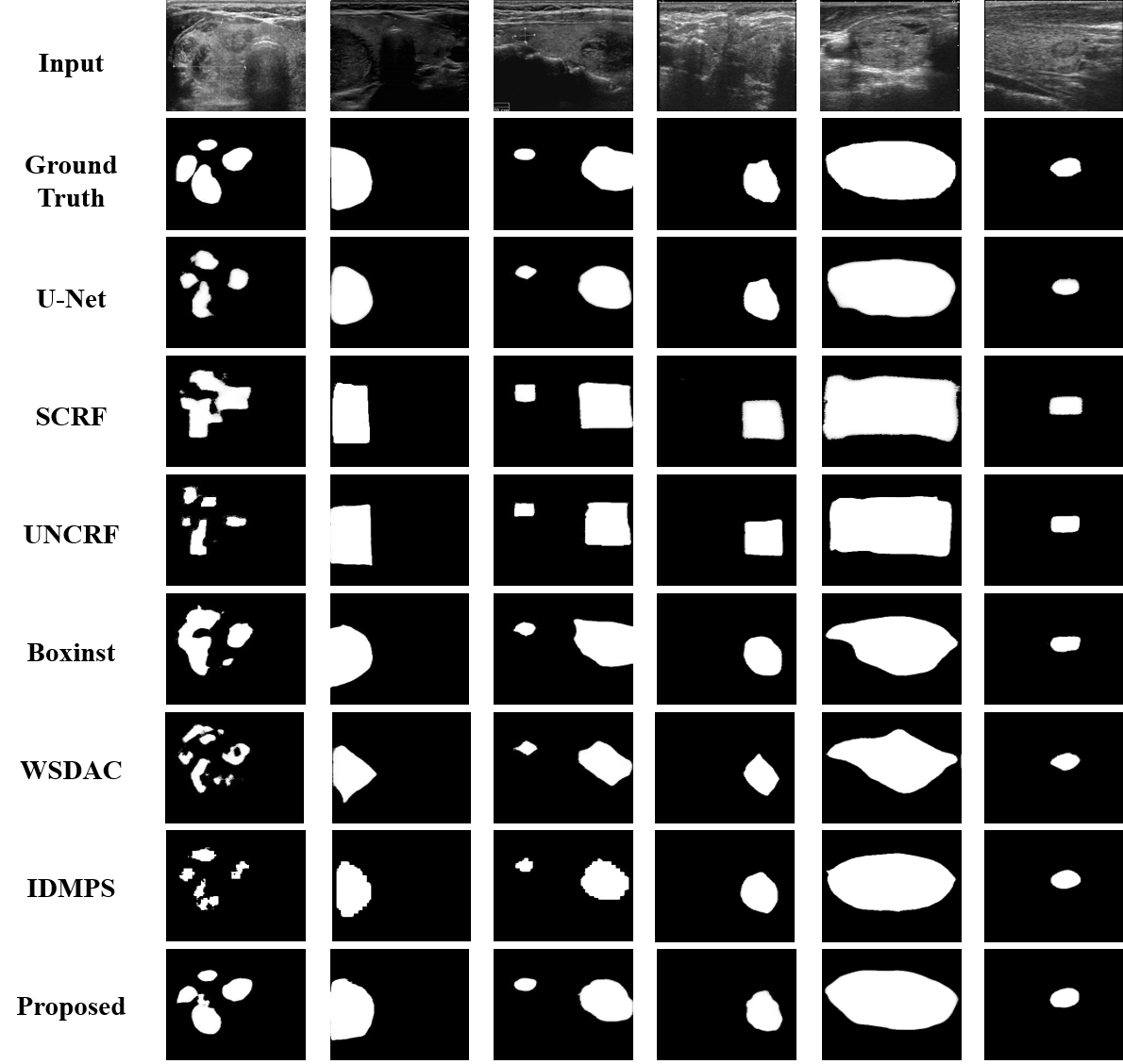}
\caption{\textbf{The segmentation visualization results of comparison experiments on TN3K and DDTI.} The three images on the left are from the TN3K dataset, while the three images on the right belong to DDTI.}
\vspace{-0.15in}
\label{fig:comparison_result}
\end{figure}

\subsubsection{Qualitative Comparison}
Fig.~\ref{fig:comparison_result} presents several example segmentation results of different methods. The SCRF and UNCRF algorithms exhibit overfitting due to incorrect classification based on box labels. The BoxInst and WSDAC methods demonstrate underfitting of large nodules due to ineffective feature learning. The segmentation edges produced by IDMPS are not smooth, as the method fails to account for multi-scale information. In contrast, our proposed network predicts more complete thyroid nodule segmentation masks in the first column, covers the nodules more effectively in other columns, and achieves better shape fitting compared to other methods.


\subsubsection{Quantitative Results}
As shown in Table~\ref{tab:comparison_result}, the proposed method outperformed other weakly supervised networks, achieving the best performance across mIoU, DSC, and HD metrics on both the TN3K and DDTI datasets. Using the same U-Net backbone, our network delivered segmentation results comparable to fully supervised methods while relying only on the point annotation, with mIoU scores of 82.36\% on TN3K and 84.33\% on DDTI. This highlights the significant potential of $\text{PA}^+$Net in effectively leveraging sparse labels for robust feature learning.


\section{Conclusion}
In this paper, we propose a weakly-supervised segmentation algorithm refining multi-level information from the clinical point annotation, and derive multi-scale constraints to supervise network based on the box label and pure foreground/background labels. The ablation experiments and comparison experiments results indicate that our proposed network can achieve superior segmentation accuracy and edge fitting. Our framework is versatile and can be easily applied to various weakly supervised segmentation backbone models to enhance their performance.

\bibliographystyle{ieeetr}
\bibliography{main}

\begin{thebibliography}{10}

\bibitem{chi2023htunet}
J.~Chi, Z.~Li, Z.~Sun, X.~Yu, and H.~Wang, ``Hybrid transformer unet for
  thyroid segmentation from ultrasound scans,'' {\em Computers in Biology and
  Medicine}, vol.~153, p.~106453, 2023.

\bibitem{tessler2017Ti-rads}
F.~N. Tessler, W.~D. Middleton, E.~G. Grant, J.~K. Hoang, L.~L. Berland, S.~A.
  Teefey, J.~J. Cronan, M.~D. Beland, T.~S. Desser, M.~C. Frates, {\em et~al.},
  ``Acr thyroid imaging, reporting and data system (ti-rads): white paper of
  the acr ti-rads committee,'' {\em Journal of the American college of
  radiology}, vol.~14, no.~5, pp.~587--595, 2017.

\bibitem{ronneberger2015unet}
O.~Ronneberger, P.~Fischer, and T.~Brox, ``U-net: Convolutional networks for
  biomedical image segmentation,'' in {\em Medical image computing and
  computer-assisted intervention--MICCAI 2015: 18th international conference,
  Munich, Germany, October 5-9, 2015, proceedings, part III 18}, pp.~234--241,
  Springer, 2015.

\bibitem{cai2020denseunet}
S.~Cai, Y.~Tian, H.~Lui, H.~Zeng, Y.~Wu, and G.~Chen, ``Dense-unet: a novel
  multiphoton in vivo cellular image segmentation model based on a
  convolutional neural network,'' {\em Quantitative imaging in medicine and
  surgery}, vol.~10, no.~6, p.~1275, 2020.

\bibitem{tao2022cenet}
H.~Tao, C.~Xie, J.~Wang, and Z.~Xin, ``Cenet: A channel-enhanced spatiotemporal
  network with sufficient supervision information for recognizing industrial
  smoke emissions,'' {\em IEEE internet of things journal}, vol.~9, no.~19,
  pp.~18749--18759, 2022.

\bibitem{zhang2020scrf}
N.~Zhang, S.~Francis, R.~A. Malik, and X.~Chen, ``A spatially constrained deep
  convolutional neural network for nerve fiber segmentation in corneal confocal
  microscopic images using inaccurate annotations,'' in {\em 2020 IEEE 17th
  International Symposium on Biomedical Imaging (ISBI)}, pp.~456--460, IEEE,
  2020.

\bibitem{mahani2022uncrf}
G.~K. Mahani, R.~Li, N.~Evangelou, S.~Sotiropolous, P.~S. Morgan, A.~P. French,
  and X.~Chen, ``Bounding box based weakly supervised deep convolutional neural
  network for medical image segmentation using an uncertainty guided and
  spatially constrained loss,'' in {\em 2022 IEEE 19th International Symposium
  on Biomedical Imaging (ISBI)}, pp.~1--5, IEEE, 2022.

\bibitem{tian2021BoxInst}
Z.~Tian, C.~Shen, X.~Wang, and H.~Chen, ``Boxinst: High-performance instance
  segmentation with box annotations,'' in {\em Proceedings of the IEEE/CVF
  Conference on Computer Vision and Pattern Recognition}, pp.~5443--5452, 2021.

\bibitem{li2023wsdac}
Z.~Li, S.~Zhou, C.~Chang, Y.~Wang, and Y.~Guo, ``A weakly supervised deep
  active contour model for nodule segmentation in thyroid ultrasound images,''
  {\em Pattern Recognition Letters}, vol.~165, pp.~128--137, 2023.

\bibitem{zhao2024IDMPS}
X.~Zhao, Z.~Li, X.~Luo, P.~Li, P.~Huang, J.~Zhu, Y.~Liu, J.~Zhu, M.~Yang,
  S.~Chang, {\em et~al.}, ``Ultrasound nodule segmentation using asymmetric
  learning with simple clinical annotation,'' {\em IEEE Transactions on
  Circuits and Systems for Video Technology}, 2024.

\bibitem{gong2021tn3k}
H.~Gong, G.~Chen, R.~Wang, X.~Xie, M.~Mao, Y.~Yu, F.~Chen, and G.~Li,
  ``Multi-task learning for thyroid nodule segmentation with thyroid region
  prior,'' in {\em 2021 IEEE 18th international symposium on biomedical imaging
  (ISBI)}, pp.~257--261, IEEE, 2021.

\bibitem{pedraza2015DDTI}
L.~Pedraza, C.~Vargas, F.~Narv{\'a}ez, O.~Dur{\'a}n, E.~Mu{\~n}oz, and
  E.~Romero, ``An open access thyroid ultrasound image database,'' in {\em 10th
  International symposium on medical information processing and analysis},
  vol.~9287, pp.~188--193, SPIE, 2015.

\end{thebibliography}

\end{document}